\documentclass[a4paper,12pt]{article}

\usepackage{amsmath,amsfonts,amssymb,mathtools}
\usepackage{mathbbol,mathrsfs,dsfont,bm}
\usepackage{latexsym,wasysym}
\usepackage{cancel,slashed,units}

\usepackage{hepnicenames,feynmp-auto}

\usepackage{graphicx,rotating}
\usepackage[table,xcdraw,dvipsnames]{xcolor}

\usepackage{booktabs,multirow,multicol}
\usepackage{enumerate,expdlist}

\usepackage{soul,comment,xstring}
\usepackage[normalem]{ulem}

\usepackage[nosort]{cite}
\usepackage{hyperref}

\hypersetup{
  colorlinks=true,
  bookmarksopen=true,
  bookmarksnumbered=true,
  pdfstartview=FitH,
  linkcolor=blue!60!black,
  urlcolor=red!60!black,
  citecolor=green!50!black
}

\allowdisplaybreaks

\newcommand{\mio}[1]{}
\newcommand{\gfootnote}[1]{}

\newcommand{\beq}{\begin{equation}}
\newcommand{\eeq}{\end{equation}}
\newcommand{\be}{\begin{equation}}
\newcommand{\ee}{\end{equation}}

\newcommand{\beqa}{\begin{equation}\begin{aligned}}
\newcommand{\eeqa}{\end{aligned}\end{equation}}

\newcommand{\bea}{\begin{eqnarray}}
\newcommand{\eea}{\end{eqnarray}}

\newcommand{\bpm}{\begin{pmatrix}}
\newcommand{\epm}{\end{pmatrix}}

\newcommand{\sfrac}[2]{#1/#2}

\newcommand{\diag}{\,\operatorname{diag}}

\newcommand{\fig}[1]{~\ref{fig:#1}}
\newcommand{\eq}[1]{~\eqref{eq:#1}}

\newcommand{\hhref}[1]{%
  \href{https://arxiv.org/abs/#1}{arXiv:#1}%
}
\newcommand{\hhrefq}[1]{%
  \IfSubStr{#1}{:}
    {\href{https://inspirehep.net/search?q=#1}{InSpire:#1}}
    {\hhref{#1}}%
}
\newcommand{\hhrefqq}[1]{%
  \IfBeginWith{#1}{10.}
    {\href{https://doi.org/#1}{doi:#1}}
    {\hhrefq{#1}}%
}

\newcommand{\doi}[1]{\href{https://doi.org/#1}{[link]}}

\makeatletter

\def\art{\@ifnextchar[{\eart}{\oart}}
\def\oart#1#2#3#4#5{%
  \textrm{#1}, \textit{#2 \textbf{#3}} \textrm{(#5) #4}%
}
\def\eart[#1]#2#3#4#5#6{%
  \oart{#2}{#3}{#4}{#5}{#6} (\textit{#1})%
}

\def\article{\@ifnextchar[{\earticle}{\oarticle}}
\def\oarticle#1#2#3#4#5#6{%
  \textrm{#1}, \textit{`#6'}, \textrm{#2 #3 (#5) #4}%
}
\def\earticle[#1]#2#3#4#5#6#7{%
  \oarticle{#2}{#3}{#4}{#5}{#6}{#7} [\hhrefq{#1}]%
}

\def\hepart[#1]#2{\textrm{#2}, {\slshape #1}}
\def\heparticle[#1]#2#3{%
  #2, \textit{`#3'} [\hhrefq{#1}]%
}

\renewenvironment{thebibliography}[1]
  {%
    \begin{multicols}{2}[\section*{\refname}]%
    \@mkboth{\MakeUppercase\refname}{\MakeUppercase\refname}%
    \list{\@biblabel{\@arabic\c@enumiv}}%
      {%
        \settowidth\labelwidth{\@biblabel{#1}}%
        \leftmargin\labelwidth
        \advance\leftmargin\labelsep
        \@openbib@code
        \usecounter{enumiv}%
        \let\p@enumiv\@empty
        \renewcommand\theenumiv{\@arabic\c@enumiv}%
      }%
    \sloppy
    \clubpenalty4000
    \@clubpenalty\clubpenalty
    \widowpenalty4000
    \sfcode`\.\@m
  }
  {%
    \def\@noitemerr{%
      \@latex@warning{Empty `thebibliography' environment}%
    }%
    \endlist
    \end{multicols}%
  }

\makeatother

\newcounter{alphaequation}[equation]
\renewcommand{\thealphaequation}{%
  \theequation\hbox to 0.6em{\hfil\alph{alphaequation}\hfil}%
}

\makeatletter

\newcommand{\eqnsystem}[1]{%
  \renewcommand{\@eqnnum}{\textrm{(\thealphaequation)}}%
  \renewcommand{\@@eqncr}{%
    \let\@tempa\relax
    \ifcase\@eqcnt
      \def\@tempa{& & &}%
    \or
      \def\@tempa{& &}%
    \or
      \def\@tempa{&}%
    \fi
    \@tempa
    \if@eqnsw
      \@eqnnum\refstepcounter{alphaequation}%
    \fi
    \global\@eqnswtrue
    \global\@eqcnt=0
    \cr
  }%
  \refstepcounter{equation}%
  \let\@currentlabel\theequation
  \def\@tempb{#1}%
  \ifx\@tempb\empty\else\label{#1}\fi
  \refstepcounter{alphaequation}%
  \let\@currentlabel\thealphaequation
  \global\@eqnswtrue
  \global\@eqcnt=0
  \tabskip\@centering
  \let\\=\@eqncr
  $$\halign to \displaywidth\bgroup
    \@eqnsel\hskip\@centering
    $\displaystyle\tabskip\z@{##}$%
    &\global\@eqcnt\@ne
    \hskip2\arraycolsep\hfil${##}$\hfil
    &\global\@eqcnt\tw@
    \hskip2\arraycolsep
    $\displaystyle\tabskip\z@{##}$\hfil
    \tabskip\@centering
    &\llap{##}\tabskip\z@\cr
}

\def\endeqnsystem{%
  \@@eqncr\egroup$$\global\@ignoretrue
}

\makeatother

\begin{document}

\begin{center}
{\bfseries\LARGE\color{red!50!black} A Texture of Leptons 
for the \\ Koide and Alikhanov relations}\\[2ex]
{\bfseries Alessandro Strumia}\\[1ex]
{\itshape Dipartimento di Fisica, Universit\`a di Pisa, Italia}\\[2ex]
{\bfseries\color{blue!40!black} Abstract}
\begin{quote} \color{blue!40!black}
Koide and Alikhanov proposed intriguing empirical relations among
charged-lepton masses and neutrino mass splittings, respectively.
We present mass-matrix textures that realise these relations and
can jointly account for the observed lepton mixing angles, while
yielding predictive correlations among them and the neutrino CP phases.
\end{quote}
\end{center}

\section{Introduction}
Alikhanov and Koide noticed that measured neutrino and charged lepton masses
are compatible with the empirical relations
\begin{equation}\label{eq:Alikhanov}
 \frac{\sqrt{\Delta m^2_{31}}+\sqrt{\Delta m^2_{21}}}
 {\sqrt{\Delta m^2_{31}}-\sqrt{\Delta m^2_{21}}}=\sqrt2~
 \hbox{\cite{2601.18781,2610.06738}},\qquad
  \frac{m_e+m_\mu+m_\tau}{(\sqrt{m_e}+\sqrt{m_\mu}+\sqrt{m_\tau})^2}
 =\frac23~
 \hbox{\cite{Koide:1983qe,hep-ph/9402242,hep-ph/0005137,hep-ph/0509214,hep-ph/0605074,0706.2534,0708.3913,hep-ph/0602134,0812.2103,1203.2028}} .
\end{equation}
Here $\Delta m^2_{ij}=m_i^2-m_j^2$ and neutrino normal ordering is assumed. 
With the latest charged-lepton pole masses~\cite{ParticleDataGroup:2024cfk},
the Koide relation holds within $1\sigma$, with a relative accuracy of order $10^{-5}$.
QED radiative corrections degrade this agreement when the relation is expressed
in terms of running masses, yielding a relative deviation of about $0.2\%$
at the electroweak scale~\cite{hep-ph/0602134}.
The Alikhanov relation is also compatible with data within $1\sigma$,
although currently tested only at a relative precision of about
$0.3\%$~\cite{2610.06738}.
Defining $Q_N (z_i)= \sum z_i^2/(\sum z_i)^2$ for $z_{1,\ldots ,N}\ge 0$,
Alikhanov corresponds to $Q_2(\sqrt{\Delta m^2})=3/4$~\cite{2601.18781},
Koide to $Q_3(\sqrt{m_\ell})=2/3$. 
They can be written in a common form  as $Q_N =(N+1)/2N$, 
the average between the minimal degenerate value $1/N$ and the maximal hierarchical value $1$.

\section{Neutrino masses: the Alikhanov relation}
Inspired by the golden-ratio texture~\cite{Kajiyama:2007gx}, we consider its silver-ratio analogue
\begin{equation}
M_\nu=m_0 A_1,\qquad A_1=\begin{pmatrix}
0&0&0\\
0&0&1\\
0&1&2
\end{pmatrix}.
\end{equation}
The resulting neutrino masses 
\begin{equation}
(m_1,m_2,m_3)=m_0 (0,\sqrt2-1,\sqrt2+1)
\end{equation}
satisfy the Alikhanov relation.
$A_1$ is diagonalized in the 23 sector by the `silver angle' $\theta=\sfrac{\pi}{8}$,
and the Alikhanov relation is $\sqrt{m_2/m_3} = \tan\theta$.
A related simple texture that gives the same  $  (m_1,m_2,m_3)$ is
\begin{equation}
 M_\nu=\frac{m_0}{\sqrt{2}} A_2,\qquad
 A_2=\begin{pmatrix}1&1&1\\1&1&1\\1&1&2\end{pmatrix}.
 \label{eq:A}
\end{equation}
Although $A_1$ and $A_2$ yield the same physical masses, with $m_1=0$ and $m_{2,3}>0$, their second eigenvalues have opposite signs, corresponding to a relative Majorana phase of $\pi$.
A two-singlet seesaw produces this texture from
\begin{equation}
 m_D=v Y,\qquad Y=\begin{pmatrix}1&0\\1&0\\1&1\end{pmatrix}= (a\ b),\qquad
 M_R=M\mathbf1_2.
\end{equation}
In these notations $A_2= Y Y^T=aa^T+bb^T$ with $a=(1,1,1)^T$ and $b=(0,0,1)^T$. 
The associated Gram matrix 
\begin{equation}
 G =
Y^T Y= \begin{pmatrix}a\cdot a&a\cdot b\\b\cdot a&b\cdot b\end{pmatrix}
 =\begin{pmatrix}3&1\\1&1\end{pmatrix}.
\end{equation}
has the same nonzero eigenvalues as $A_2$.
In terms of $G$ the Alikhanov condition is $\det G/(\operatorname{tr}G)^2=m_2 m_3/(m_2+m_3)^2=1/8$.
A third see-saw-derivable texture that gives the same masses is obtained flipping the off-diagonal signs of $G$,
rescaling it, and finding a simple texture
\beq M_\nu=\frac{m_0}{3\sqrt2}A_3,\qquad
A _3=\begin{pmatrix}
2&1&1\\
1&5&5\\
1&5&5
\end{pmatrix}=aa^T+bb^T = Y Y^T
\qquad
Y=(a\ b)=\begin{pmatrix}
1&1\\
2&-1\\
2&-1
\end{pmatrix}.
\eeq
The silver angle diagonalizes these $G$, while the flavour mixing angles also depend on the embedding specified by \(Y\).

\medskip

We next compute the neutrino mixing angles.
The physical PMNS matrix is usually parameterized as
\begin{equation}
 V
 =R_{23}(\theta_{23})\,
 \operatorname{diag}(1,1,e^{i\delta})\,
 R_{13}(\theta_{13})\,
 \operatorname{diag}(1,1,e^{-i\delta})\,
 R_{12}(\theta_{12})\,
 \operatorname{diag}(e^{i\alpha_1/2},e^{i\alpha_2/2},1).
\end{equation}
The measured mixing angles are~\cite{NuFIT}
\beq
 \sin^2\theta_{12}=0.309\pm0.0067,\qquad
 \sin^2\theta_{23}=0.470\pm0.016,\qquad
 \sin^2\theta_{13}=0.0225\pm0.0006
\eeq
while the Majorana phases $\alpha_{1,2}$ are unknown, and $\delta$ is largely unknown.
An extra solution exists for $\theta_{23}$.
The PMNS matrix is the mismatch between the
left-handed neutrino and charged-lepton mass eigenstates.
We diagonalise them as
\begin{equation}
 V_\nu^T M_\nu V_\nu=\operatorname{diag}(m_1,m_2,m_3),
 \qquad
 M_e=V_L\,\operatorname{diag}(m_e,m_\mu,m_\tau)\,V_{R}^\dagger.
\end{equation}
The two mass-matrix textures $M_\nu $ and $M_e$ need not be aligned in the same phase convention
for the lepton doublets $L_{e,\mu,\tau}$.
We parametrise their relative alignment as
\begin{equation}\label{eq:relativePhi}
 V=V_L^\dagger \Phi^\dagger V_\nu,
 \eeq
 by introducing an extra diagonal phase matrix  $\Phi=\operatorname{diag}(e^{i\phi},e^{i\psi},1)$.
Only two relative phases are needed, since a common phase does not affect the mixing observables.
With the displayed texture entries held fixed, the two relative phases are physical parameters.
 In a common weak basis, these phases can be assigned to either the neutrino or the charged-lepton mass matrix.

\section{Charged lepton masses: the Koide relation}

The Koide relation involves square roots of masses, so a tentative interpretation writes the 
charged lepton mass matrix as the square of some other matrix,
$M_e=m_* K ^2$ with $m_*>0$.
This see-saw-like structure can be realised e.g.\ in `partial compositeness' models.
Then, the Koide relation follows if the positive matrix $K $  satisfies $\sfrac{\operatorname{tr}K ^2}{(\operatorname{tr}K )^2}=\sfrac23$~\cite{hep-ph/0509214}.

\smallskip

A first  cyclic texture that realises this condition is
\begin{equation}
 K_1 =\begin{pmatrix}
 1&e^{i\varphi}/\sqrt2&e^{-i\varphi}/\sqrt2\\
 e^{-i\varphi}/\sqrt2&1&e^{i\varphi}/\sqrt2\\
 e^{i\varphi}/\sqrt2&e^{-i\varphi}/\sqrt2&1
 \end{pmatrix}.
\end{equation}
All lepton masses are reproduced for $\varphi \approx 2/9$~\cite{hep-ph/0005137,hep-ph/0605074}.
The squares of the diagonal and off-diagonal parts of $K_1$ have the same trace.
This $M_e$ is diagonalised by the trimaximal Fourier matrix 
$(V_\ell)_{jk}=\sfrac{\omega^{(j-1)k}}{\sqrt3}$ with $ \omega=e^{2\pi i/3}$.
Pairing this $K_1 $ Koide texture with the $A_3$ texture allows one to fit
the three measured neutrino mixing angles  in terms of the two relative phases 
$\phi\approx 0.320$ and $\psi\approx-1.22$, for which
\begin{equation}
\sin^2\theta_{12}\approx 0.323,\qquad
\sin^2\theta_{23}\approx 0.467,\qquad
\sin^2\theta_{13}\approx 0.0225.
\end{equation}
The CP-violating angle in neutrino oscillations is predicted as $\cos\delta\approx-0.858 \pm 0.024$.
The best-fit  point has $\chi^2\approx 4.7$, which is good but not perfect.
This happens because the model predicts 
$ |V_{e1}|^2 \equiv \cos^2\theta_{12} \cos^2\theta_{13}= [1-\cos(\psi-\sfrac{2\pi}{3})]/3 \le 2/3$
which needs mildly increasing $\theta_{12}$ above its central value.
The Majorana phase is  $\alpha_2\approx 253^\circ$ at the quoted best-fit point,
or $107^\circ$ for its CP-conjugate solution; $\alpha_1$ is unphysical since $m_1=0$.
The Majorana mass relevant for neutrino-less double-beta decay is $m_{\beta\beta}\approx 3.6\,{\rm meV}$.
The fit shows an extra local minimum with $\chi^2=7$ at $\cos\delta \approx +0.80$.


\medskip

We next present an alternative Koide texture $K_2$ that, combined with $A_2$,
fits data without extra CP-violating phases, namely setting $\Phi =\diag(1,1,1)$.
This less nice alternative was selected  scanning among a few thousand similar possibilities:
\begin{equation}\label{eq:arrow}
 \begin{aligned}
 K_2&=\alpha\,\mathbf1_3+3(1-\alpha)\,bb^T
 +\beta\,(b\,c^T+c\,b^T)=\begin{pmatrix}
 \alpha&0&2\beta\\
 0&\alpha&-\beta\\
 2\beta&-\beta&3-2\alpha
 \end{pmatrix}.
 \end{aligned}
\end{equation}
Here $ b=(0,0,1)^T$ is the same vector that enters $A_2=aa^T+bb^T$, while $c=(2,-1,0)^T$ is an extra vector.
The $-1$ in $c$ is an arbitrary phase choice when computing the charged-lepton spectrum only.
The parameter $\beta$ needs to be fixed such that the Koide relation is satisfied
\begin{equation}
\frac{\operatorname{tr}K_2^2}{(\operatorname{tr}K_2)^2}
=\frac{3+6(1-\alpha)^2+10\beta^2}{9}
=\frac23.
\end{equation}
So $K_2$ is a Koide-motivated quadratic texture that fits charged lepton masses.
In addition to the Koide relation, it arbitrarily assumes:
equal first two diagonal entries, a vanishing $12$ entry, and a $2:1$ ratio of the off-diagonal magnitudes. 
The eigenvalues of $K_2$ are
\begin{equation}
 \lambda_\mu=\alpha,\qquad
 \lambda_{e,\tau}=\frac{3-\alpha\mp\sqrt{9(\alpha-1)^2+20\beta^2}}2,
 \qquad m_\ell=m_*\lambda_\ell^2.
\end{equation}
Using $m_e=0.51099895\,\mathrm{MeV}$ and $m_\mu=105.6583755\,\mathrm{MeV}$, 
Koide predicts $m_\tau\approx 1776.97\,\mathrm{MeV}$ and needs
 $\alpha\approx 0.5802$, $\beta\approx 0.4408$.
Thus the charged-lepton spectrum fixes $\alpha$, $\beta$ and $m_*$.

\begin{table}[t]
\begin{center}
\begin{tabular}{c|cc|ccccc|c}
point & $\psi$ & $\phi$ & $\sin^2\theta_{12}$ & $\sin^2\theta_{13}$ & $\sin^2\theta_{23}$ & $\delta$ & $m_{\beta\beta}$& $\chi^2$\\
\midrule
0 & $0$ & $0$ & $0.29315$ & $0.02062$ & $0.45947$ & $180^\circ$ & $3.5$ meV & $16$\\
1 & $\pi/8$ & $0$ & $0.31596$ & $0.02219$ & $0.44464$ & $221^\circ$ & $3.1$ meV & $4.1$\\
2 & $\pi/4$ & $\pi/6$ & $0.30248$ & $0.02254$ & $0.45340$ & $149^\circ$ & $3.3$ meV &  $2.1$\\
\end{tabular}
\caption{\em\label{tab:fit} Considering the $A_2+K_2$ texture
we show, for the given values of the extra phases $\psi$ and $\phi$,
the predicted neutrino mixings, the CP phase in neutrino oscillations $\delta$,
the effective Majorana mass $m_{\beta\beta}$ and the $\chi^2$ values.
Its minimum is $\min \chi^2\approx 1.7$. The three points are illustrated in fig.\fig{AliKoidePredictions}.}
\end{center}
\end{table}

\begin{figure}[t]
$$\includegraphics[width=0.9\textwidth]{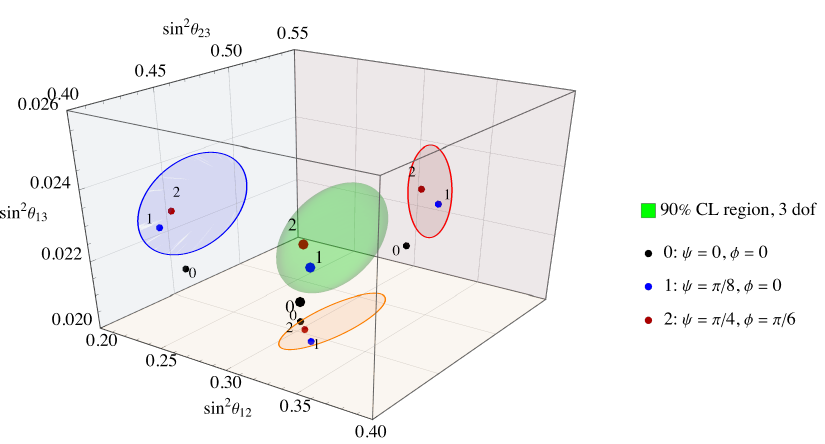}$$
\begin{center}
\caption{\em\label{fig:AliKoidePredictions} Predicted values of the neutrino mixing angles in the
$A_2+K_2$ texture (point 0), and adding the extra values of the CP phases in table~\ref{tab:fit}.
Projections along the three colored planes are shown.}
\end{center}
\end{figure}

The corresponding real charged-lepton diagonalization matrix, ordered as $e,\mu,\tau$, is
\begin{equation}\label{eq:Ue}
 V_L=\begin{pmatrix}
 2C/\sqrt5&1/\sqrt5&2S/\sqrt5\\
 -C/\sqrt5&2/\sqrt5&-S/\sqrt5\\
 -S&0&C
 \end{pmatrix}.
\end{equation}
We defined $C=\cos\chi>0$, $S=\sin\chi>0$, with
\begin{equation}\label{eq:chargedangle}
 C^2=\frac{\lambda_\tau-\alpha}{\lambda_\tau-\lambda_e}
 =\frac{\sqrt{m_\tau}-\sqrt{m_\mu}}{\sqrt{m_\tau}-\sqrt{m_e}}
 \approx0.7692,\qquad S^2=1-C^2.
\end{equation}
$V_L$ depends on the charged-lepton hierarchy through $\chi\approx 28.7^\circ$.
Combining $K_2$ with $A_2$ assuming no extra phases, $\phi=\psi=0$ leaves no extra free parameters.
Yet this combination is compatible with all three measured lepton angles in a reasonably acceptable way, see table~\ref{tab:fit} and fig.\fig{AliKoidePredictions}.
The $\theta_{13}$ angle is $3\sigma$ below the measured value.
The real $K_2+A_2$ combination predicts no CP violation.

\medskip

Allowing for generic CP-violating phases $\phi,\psi$, the $K_2, A_2$ texture still predicts a relation between charged-lepton masses and mixing angles:
\begin{equation}\label{eq:massmixing}
 |V_{e1}|^2=2C^2|V_{\mu3}|^2,\qquad\hbox{i.e.}\qquad
 {\sin^2\theta_{23}
 =\frac{\sqrt{m_\tau}-\sqrt{m_e}}
 {2(\sqrt{m_\tau}-\sqrt{m_\mu})}\cos^2\theta_{12}
 \approx0.650\cos^2\theta_{12}.}
\end{equation}
Furthermore, table~\ref{tab:fit} lists values of the Majorana mass relevant for neutrino-less double-beta decay
\begin{equation}\label{eq:mbb}
 m_{\beta\beta}=
 \frac{m_0}{\sqrt2}\bigg|\bigg [\frac{C(2e^{-i\phi}-e^{-i\psi})}{\sqrt5}-S\bigg]^2+S^2\bigg|.
\end{equation}
The Alikhanov relation eq.\eq{Alikhanov} survives adding to $A_{1,2,3}$ 
the unit matrix with imaginary coefficient. This would modify $m_{\beta\beta}$.

\medskip

In conclusion, these textures do not constitute a flavour model: 
a dynamical explanation of their structure and relative alignment remains to be found.
In particular, the textures realising the Koide relation are contrived and lack a compelling symmetry interpretation.
Moreover, their numerical success was achieved 
after exploring many possibilities.

{\small
\paragraph{Acknowledgements.}
I thank ChatGPT and Claude for exploring many textures.
}


\footnotesize

\begin{thebibliography}{nnn}\bibitem{2601.18781} I.~Alikhanov, \textit{Another relation among the neutrino mass-squared differences?}, \hhref{2601.18781}.


\bibitem{2610.06738}
\heparticle[2610.06738]{I. Alikhanov}{Novel dependence between neutrino mass splittings strongly supported by initial JUNO results}.


\bibitem{Koide:1983qe}
\article{Y. Koide}{Phys.Rev.D}{28}{252}{1983}
{\href{https://doi.org/10.1103/PhysRevD.28.252}{A New View of Quark and Lepton Mass Hierarchy}}.


\bibitem{hep-ph/9402242}
\heparticle[hep-ph/9402242]{R. Foot}{A Note on Koide's lepton mass relation}.


\bibitem{hep-ph/0005137}
\heparticle[hep-ph/0005137]{Y. Koide}{Quark and lepton mass matrices with a cyclic permutation invariant form}.


\bibitem{hep-ph/0509214}
\article[hep-ph/0509214]{Y. Koide}{Phys.Rev.D}{73}{057901}{2006}
{\href{https://doi.org/10.1103/PhysRevD.73.057901}{Permutation symmetry S(3) and VEV structure of flavor-triplet Higgs scalars}}.


\bibitem{hep-ph/0605074}
\article[hep-ph/0605074]{Y. Koide}{J.Phys.G}{34}{1653}{2007}
{\href{https://doi.org/10.1088/0954-3899/34/7/006}{Tribimaximal Neutrino Mixing and a Relation Between Neutrino- and Charged Lepton-Mass Spectra}}.


\bibitem{0706.2534}
\article[0706.2534]{Y. Koide}{Int.J.Mod.Phys.E}{16}{1417}{2007}
{\href{https://doi.org/10.1142/S0218301307006770}{Charged Lepton Mass Formula: Development and Prospect}}.


\bibitem{0708.3913}
\article[0708.3913]{N. Haba, Y. Koide}{Phys.Lett.B}{659}{260}{2008}
{\href{https://doi.org/10.1016/j.physletb.2007.10.082}{New Origin of a Bilinear Mass Matrix Form}}.


\bibitem{hep-ph/0602134}
\article[hep-ph/0602134]{Z.-Z. Xing, H. Zhang}{Phys.Lett.B}{635}{107}{2006}
{\href{https://doi.org/10.1016/j.physletb.2006.02.051}{On the Koide-like relations for the running masses of charged leptons, neutrinos and quarks}}.


\bibitem{0812.2103}
\article[0812.2103]{Y. Sumino}{JHEP}{05}{075}{2009}
{\href{https://doi.org/10.1088/1126-6708/2009/05/075}{Family Gauge Symmetry as an Origin of Koide's Mass Formula and Charged Lepton Spectrum}}.


\bibitem{1203.2028}
\article[1203.2028]{Y. Koide, T. Yamashita}{Phys.Lett.B}{711}{384}{2012}
{\href{https://doi.org/10.1016/j.physletb.2012.04.028}{Family Gauge Bosons with an Inverted Mass Hierarchy}}.


\bibitem{ParticleDataGroup:2024cfk}
\article{{\sc Particle Data Group} Collaboration}{Phys.Rev.D}{110}{030001}{2024}
{\href{https://doi.org/10.1103/PhysRevD.110.030001}{Review of particle physics}}.


\bibitem{Kajiyama:2007gx}
\article[0705.4559]{Y. Kajiyama, M. Raidal, A. Strumia}{Phys.Rev.D}{76}{117301}{2007}
{\href{https://doi.org/10.1103/PhysRevD.76.117301}{The Golden ratio prediction for the solar neutrino mixing}}.


\bibitem{NuFIT} \article[2601.09791]{I. Esteban, M.C. Gonzalez-Garcia, M. Maltoni, I. Martinez-Soler, J.P. Pinheiro, T. Schwetz}{JHEP}{04}{089}{2026}
{\href{https://doi.org/10.1007/JHEP04(2026)089}{Lessons from the first JUNO results}}.



%


\end{thebibliography}
\end{document}